\documentclass[a4paper, 11pt]{article}
\usepackage{geometry}
\usepackage[utf8]{inputenc}
\usepackage[english]{babel}

\usepackage{authblk}

\usepackage{dsfont} 
\usepackage{amsmath,amssymb,amsthm}
\usepackage[justification=centering]{caption}
\usepackage[colorlinks=true, citecolor=red, linkcolor=blue]{hyperref}
\usepackage[nameinlink, capitalize]{cleveref}

\usepackage[colorinlistoftodos]{todonotes}
\usepackage{varwidth}

\usepackage{color}
\usepackage{float}
\usepackage[super]{nth}
\usepackage{booktabs}
\usepackage{rotating}
\usepackage{verbatim}
\usepackage{multirow}
\usepackage{physics}
\usepackage{mathtools}
\usepackage{subcaption}
\usepackage{bm}
\usepackage[toc,page]{appendix}
\usepackage[shortlabels]{enumitem}
\usepackage{nomencl}

\newcounter{mnotecount}[section]

\theoremstyle{plain}
\newtheorem{theorem}{Theorem}

\newtheorem{assumption}{Assumption}

\theoremstyle{definition}

\newtheorem{remark}{Remark}

\title{Nonlinear PDE models in semi-relativistic quantum physics}
\author[a]{Jakob Möller}
\author[a]{Norbert J. Mauser}

\affil[a]{Research Platform MMM "Mathematics-Magnetism-Materials" c/o Fak. Mathematik, Univ. Wien, Oskar-Morgenstern-Platz 1, 1090 Vienna, Austria}

\begin{document}

\maketitle

\begin{abstract}
    We present the self-consistent Pauli equation, a semi-relativistic model for charged spin-$1/2$-particles with self-interaction with the electromagnetic field. The Pauli equation arises as the $O(1/c)$ approximation of the relativistic Dirac equation. The fully relativistic self-consistent model is the Dirac-Maxwell equation  where the description of spin and the magnetic field arises naturally. In the non-relativistic setting the correct self-consistent equation is the Schrödinger-Poisson equation which does not describe spin and the magnetic field and where the self-interaction is with the electric field only.
    
    The Schrödinger-Poisson equation also arises as the mean field limit of the $N$-body Schrödinger equation with Coulomb interaction. We propose that the Pauli-Poisson equation arises as the mean field limit $N \rightarrow \infty$ of the linear $N$-body Pauli equation with Coulomb interaction where one has to pay extra attention to the fermionic nature of the Pauli equation.

    We present the semiclassical limit of the Pauli-Poisson equation by the Wigner method to the Vlasov equation with Lorentz force coupled to the Poisson equation which is also consistent with the hierarchy in $1/c$ of the self-consistent Vlasov equation. This is a non-trivial extension of the groundbreaking works by Lions \& Paul and Markowich \& Mauser, where we need methods like magnetic Lieb-Thirring estimates.
\end{abstract}

\section{Model hierarchy}

Relativistic quantum mechanics is an immensely successful theory giving a correct description of the behavior of particles on the atomic scale moving at high velocities compared to the speed of light. On the other hand, non-relativistic quantum mechanics is centered around the Schrödinger equation which is insufficient when relativistic effects such as spin and the magnetic field arise. In relativistic quantum mechanics the description of spin arises naturally in the Dirac equation which is the correct equation for particles with spin $1/2$, i.e. fermions. Semi-relativistic quantum mechanics is the theory that keeps relativistic corrections up to order $O(1/c)$ and it has been discovered by Wolfgang Pauli in 1927 that the correct semi-relativistic equation describing charged spin-1/2-particles in the electromagnetic field is the Pauli equation which describes spin and magnetic field through the Stern-Gerlach term.

Since charged particles emit radiation the effect of self-interaction of a charged particle with the electromagnetic field it generates cannot be neglected in a (semi-)relativistic setting. In the fully relativistic regime this effect is described by the \emph{Dirac-Maxwell equation} for spin-$1/2$-particles. Since Maxwell's equations are relativistic and Lorentz invariant it is natural to couple them self-consistently to the Dirac equation. In the non-relativistic regime
the \emph{Schrödinger-Poisson equation} offers a description of the self-interaction with the electric field which is given by a Poisson equation for $V^{\hbar}$ with the particle density $\rho^{\hbar}$ as a source term. The magnetic field, being a relativistic effect, does not self-interact at the non-relativistic level and neither does spin which is naturally coupled to the magnetic field.
In the semi-relativistic regime of $O(1/c)$ the correct self-consistent description is given by the \emph{Pauli-Poisswell equation}. Here, Maxwell's equations for the magnetic vector potential $A^{\hbar,c}$ and the electric scalar potential $V^{\hbar,c}$ are replaced by magnetostatic Poisson equations (cf. \cite{besse2007numerical}) with the current density $J^{\hbar,c}$ and the particle density $\rho^{\hbar,c}$ as source terms. This is appropriate when the typical velocity of the system is small compared to the speed of light. The $O(1/c^2)$ approximation of the Dirac equation is the \emph{Pauli-Darwin equation} \cite{itzykson2012quantum, mauser1999rigorous, mauser2000semi}.

We have the following diagram for the hierarchies of self-consistent models in relativistic quantum mechanics. The horiztontal resp. vertical arrows indicate the semiclassical ($\hbar \rightarrow 0$), resp. non-relativistic ($c \rightarrow \infty$) limits.  \\

{\begin{tabular}{ccc}
    {\fbox{\textbf{Dirac-Maxwell}}}
    & {$\stackrel{\hbar \rightarrow 0}{\longrightarrow}$}
    & {\fbox{\begin{varwidth}{\textwidth}\centering\bf rel. Vlasov-Lorentz - Maxwell \end{varwidth}}} \\[3mm]
  {$\downarrow$}
    &
    & {$\downarrow$} \\[3mm]

{\fbox{\bf Pauli-Darwin $O(1/c^2)$}}
    & {$\stackrel{\hbar \rightarrow 0}{\longrightarrow}$}
    & {\fbox{\begin{varwidth}{\textwidth}\centering\bf rel. Vlasov-Lorentz - Darwin $O(1/c^2)$\end{varwidth}}} \\[3mm]
   {\fbox{\bf Pauli-Poisswell $O(1/c)$}}
    & {$\stackrel{\hbar \rightarrow 0}{\longrightarrow}$}
    & {\fbox{\begin{varwidth}{\textwidth}\centering\bf Vlasov-Lorentz - Poisswell $O(1/c)$\end{varwidth}}} \\[3mm]

{$\downarrow$}
    &
    & {$\downarrow$} \\[3mm]

      {\fbox{\bf Pauli-Poisson}}
    & {$\stackrel{\hbar \rightarrow 0}{\longrightarrow}$}
    & {\fbox{\begin{varwidth}{\textwidth}\centering\bf Vlasov-Lorentz  - Poisson \end{varwidth}}} \\[3mm]
{\fbox{\textbf{magn. Schrödinger-Maxwell}}}
    & {$\stackrel{\hbar \rightarrow 0}{\longrightarrow}$}
    & {\fbox{\begin{varwidth}{\textwidth}\centering\bf non-rel. Vlasov-Lorentz - Maxwell \end{varwidth}}} \\[3mm]
    {\fbox{\textbf{magn. Schrödinger-Poisson}}}
    & {$\stackrel{\hbar \rightarrow 0}{\longrightarrow}$}
    & {\fbox{\begin{varwidth}{\textwidth}\centering\bf Vlasov-Lorentz - Poisson \end{varwidth}}} \\[3mm]
  {$\downarrow$}
    &
    & {$\downarrow$} \\[3mm]
  {${c \rightarrow \infty}$}
    &
    & {${c \rightarrow \infty}$} \\[3mm]
  $\downarrow$
    &
    & $\downarrow$ \\[3mm]
  {\fbox{\bf Schr\"odinger-Poisson}}
    & {$\stackrel{\hbar \rightarrow 0}{\longrightarrow} $} 
    & {\fbox{\begin{varwidth}{\textwidth}\centering\bf Vlasov-Poisson \end{varwidth}}}
    \\
\end{tabular}}\\
\bigskip

\subsection{Pauli-Poisswell equation: A consistent $O(1/c)$ model}

In the fully self-consistent semi-relativistic model where a magnetostatic $O(1/c)$ approximation of  Maxwell's  equations is used to self-consistently describe the magnetic field, the magnetic potential $A^{\hbar,c}$ (depending on $\hbar$ and $c$) is coupled to $\Psi^{\hbar,c}$ via three Poisson type equations with the Pauli current density as source term. This yields the \textbf{Pauli-Poisswell equation} for a 2-spinor $\Psi^{\hbar,c} = (\psi_1^{\hbar,c},\psi_2^{\hbar,c})^T\in (L^2(\mathbb{R}^3,\mathbb{C}))^2$:
\begin{align}
    i\hbar\partial_t \Psi^{\hbar,c} &= -\frac{1}{2m}(\hbar \nabla-i\frac{q}{c}A^{\hbar,c})^2\Psi^{\hbar,c} + qV^{\hbar,c} \Psi^{\hbar,c} -\frac{\hbar q}{2mc} (\sigma \cdot B^{\hbar,c}) \Psi^{\hbar,c}, \label{eq:PPW_Pauli}\\
    \Delta V^{\hbar,c} &= -\rho^{\hbar,c} = -|\Psi^{\hbar,c}|^2, \\
    \Delta A^{\hbar,c} &= -\frac{1}{c}J^{\hbar,c} \label{eq:PPW_PoissonA}
\end{align}
where the \emph{Pauli current density} is given by
\begin{equation}
    J^{\hbar,c}(\Psi^{\hbar,c},A^{\hbar,c}) = \Im(\overline{\Psi^{\hbar,c}}(\hbar\nabla -i\frac{q}{c}A^{\hbar,c})\Psi^{\hbar,c}) -{\hbar}\nabla \times (\overline{\Psi^{\hbar,c}} \sigma \Psi^{\hbar,c}), \label{eq:PPW_current},
\end{equation}
with initial data
\begin{equation}
    \Psi^{\hbar,c}(x,0) = \Psi^{\hbar,c}_I(x) \in (L^2(\mathbb{R}^3))^2.
\end{equation}
Here, $|\Psi^{\hbar,c}|^2 = |\psi_1^{\hbar,c}|^2 +|\psi_2^{\hbar,c}|^2$. Spin and magnetic field are coupled by the \emph{Stern-Gerlach term} $\mathbf{\sigma} \cdot B^{\hbar,c} := \sum_{k=1}^3 \sigma_k B^{\hbar,c}_k$ where $B^{\hbar,c} = \nabla \times A^{\hbar,c}$ is the magnetic field and where the $\sigma_k$ are the Pauli matrices
\begin{align}
    \sigma_1 = \begin{pmatrix}
    0 & 1 \\ 1 & 0
    \end{pmatrix}, && 
     \sigma_2 = \begin{pmatrix}
    0 & -i \\ i & 0
    \end{pmatrix}, &&
     \sigma_3 = \begin{pmatrix}
    1 & 0 \\ 0 & -1
    \end{pmatrix}.
\end{align} 
The Pauli-Poisswell equation is the only consistent $O(1/c)$ approximation of the Dirac-Maxwell equation. It was derived in \cite{masmoudi2001selfconsistent}.  The two components of the Pauli equation describe the two spin states of a fermion, whereas the Poisson equations describe the electrodynamic self-interaction of a fast moving particle with the electromagnetic field that it generates itself due to the finite speed of light. Since $A^{\hbar,c}$ is coupled to $\Psi^{\hbar,c}$ we write a superscript $\hbar,c$ in order to emphasize its dependence on the semiclassical and the relativistic parameter. Compare  \eqref{eq:PPW_current} to \eqref{eq:Pauli_current} and notice that in the former the magnetic potential depends on $\hbar$ and $c$. The semiclassical limit of \eqref{eq:PPW_Pauli}-\eqref{eq:PPW_PoissonA} to the Vlasov equation with Lorentz force coupled to the Poisswell equations (\emph{Vlasov-Poisswell equation}) by the Wigner method is to be published in \cite{MaMo23}. The numerics of the Vlasov-Poisswell equation were discussed in \cite{besse2007numerical}. The existence of classical solutions was discussed in \cite{seehafer2009local}. The classical limit $c\rightarrow \infty, \hbar \rightarrow 0$ of the Dirac-Maxwell equation to the Vlasov-Poisson equation was proven in \cite{mauser2007convergence} where the authors first perform the non-relativistic limit to the Schrödinger-Poisson equation and then the semiclassical limit to the Vlasov-Poisson equation.  The semiclassical limit of the Dirac-Maxwell equation to the relativistic Vlasov-Maxwell equation is a very hard open question. We would like to mention two recent works on the regularity of weak solutions to the Vlasov-Maxwell equation by Besse \& Bechouche \cite{besse2018regularity} and Bardos, Besse \& Nguyen \cite{bardos2019onsager}. 

\subsection{Pauli-Poisson and magnetic Schrödinger-Maxwell equation}

In the situation where an external magnetic field $A$ is applied which is much stronger than the self-consistent magnetic field generated by the particle then the appropriate model is the \textbf{Pauli-Poisson equation}, given by
\begin{align} 
    i\hbar\partial_t \Psi^{\hbar,c} &= -\frac{1}{2m}(\hbar \nabla-i\frac{q}{c}A)^2\Psi^{\hbar,c} + qV^{\hbar,c} \Psi^{\hbar,c} - \frac{\hbar q}{2cm} (\sigma \cdot B) \Psi^{\hbar,c},\label{eq:PP_Pauli_unscaled}\\
    \Delta V^{\hbar,c} &= -\rho^{\hbar,c} := -|\Psi^{\hbar,c}|^2.\label{eq:PP_Poisson_unscaled}
\end{align}
with initial data
\begin{equation}
    \Psi^{\hbar,c}(x,0) = \Psi^{\hbar,c}_I(x) \in (L^2(\mathbb{R}^3))^2.
\end{equation}
and Pauli current density $J^{\hbar,c}$ given by 
\begin{equation}
 J^{\hbar,c} = \Im(\overline{\Psi^{\hbar,c}}(\hbar \nabla -i\frac{q}{c}A)\Psi^{\hbar,c}) + \hbar \nabla \times (\overline{\Psi^{\hbar,c}}\sigma \Psi^{\hbar,c}).
    \label{eq:Pauli_current}
\end{equation}More generally we may consider the \textbf{Pauli-Hartree equation}
\begin{align} 
    i\hbar\partial_t \Psi^{\hbar,c} &= -\frac{1}{2m}(\hbar \nabla-i\frac{q}{c}A)^2\Psi^{\hbar,c} + V^{\text{ext}} \Psi^{\hbar,c} - \frac{\hbar q}{2cm} (\sigma \cdot B) \Psi^{\hbar,c} + (W \ast|\Psi^{\hbar,c}|^2)\Psi^{\hbar,c},\label{eq:pauli hartree}
\end{align}
where $V^{\text{ext}}$ is an external potential and $W$ is an interaction kernel depending on $x\in \mathbb{R}^3$. In $3-d$ only the Pauli-Poisson equation corresponds to the Pauli-Hartree equation with
\begin{equation}
    W(x) = -\frac{\lambda}{|x|},
\end{equation}where $\lambda$ is a coupling constant. The Pauli-Poisson equation is related to the \emph{magnetic Schrödinger-Maxwell equation}, considered in \cite{bejenaru2009global}, the \emph{magnetic Schrödinger-Poisson equation}, considered in \cite{barbaroux2016existence, barbaroux2017well} and the \emph{magnetic Schrödinger-Hartree equation}, considered in \cite{luhrmann2012mean,michelangeli2015global}. \\

The Pauli-Poisson, magnetic Schrödinger-Maxwell and magnetic Schrödinger-Poisson equations are all inconsistent models in the small parameter $1/c$. In fact these models omit term of order $O(1/c)$ and are therefore $O(1)$ in $1/c$. The magnetic Schrödinger-Maxwell equation in Lorenz gauge is given by
\begin{equation}
i\hbar \partial_t \psi^{\hbar} = -\frac{1}{2m}(\hbar\nabla-i\frac{q}{c} A^{\hbar})^2 \psi^{\hbar} + qV^{\hbar} \psi^{\hbar},
\label{eq:Schrödinger_Maxwell}
\end{equation}
\begin{align}
\Box V^{\hbar} = 4\pi |\psi^{\hbar}|^2, 
&& \Box A^{\hbar}  = \frac{4\pi }{c} J^{\hbar} \label{eq:Maxwell2},
\end{align}
where 
\begin{equation}
  J^{\hbar} = \Im(\overline{\psi^{\hbar}}(\hbar\nabla-i\frac{q}{c} A^{\hbar})\psi^{\hbar})
  \label{eq:mSP_current}
\end{equation} 
is the current density of the magnetic Schrödinger equation and initial data
\begin{align}
    \psi^{\hbar}(x,0) = \psi^{\hbar}_I && A^{\hbar}(x,0) = a^{\hbar}_0 && \partial_tA^{\hbar}(x,0) = a^{\hbar}_1.
\end{align}
The magnetic Schrödinger-Poisson equation is given by \begin{align}
    i\hbar\partial_t \psi^{\hbar} &= -\frac{1}{2m} (\hbar\nabla-i\frac{q}{c}A)^2 \psi^{\hbar} + qV^{\hbar}\psi^{\hbar}, \label{eq:mSP_Schrödinger}\\
    \Delta V^{\hbar} &= -|\psi^{\hbar}|^2,
    \label{eq:mSP_Poisson}
\end{align}
with initial data
\begin{equation}
    \psi^{\hbar}(x,0) = \psi^{\hbar}_I(x) \in L^2(\mathbb{R}^3) \label{eq:mSP_data},
\end{equation}
The magnetic Schrödinger-Hartree equation is given by 
\begin{align}
    i\hbar\partial_t \psi^{\hbar} &= -\frac{1}{2m} (\hbar\nabla-i\frac{q}{c}A)^2 \psi^{\hbar} + V^{\text{ext}}\psi^{\hbar} + (W\ast |\psi^{\hbar}|^2)\psi^{\hbar}, \label{eq:MSP_Schrödinger}
\end{align}
with initial data
\begin{equation}
    \psi^{\hbar}(x,0) = \psi^{\hbar}_I(x) \in L^2(\mathbb{R}^3) \label{eq:MSP_data},
\end{equation}
In $\mathbb{R}^3$ with $W(x)\simeq|x|^{-1}$ we obtain the magnetic Schrödinger-Poisson equation \eqref{eq:mSP_Schrödinger}-\eqref{eq:mSP_data} from the magnetic Schrödinger-Hartree equation. Here we use a lower case $\psi^{\hbar}$ to denote a \emph{scalar} wave function.
Compare \eqref{eq:mSP_current} to \eqref{eq:Pauli_current} where we have an additional divergence-free term due to the spin coupling which is not present in \eqref{eq:mSP_current}. In \cite{barbaroux2016existence, barbaroux2017well} the global wellposedness for bounded external potentials was shown. In \cite{luhrmann2012mean} the mean field limit of the $N$-body magnetic Schrödinger equation to the 1-body magnetic Schrödinger-Hartree equation was proved and in \cite{michelangeli2015global} the global wellposedness of the magnetic Schrödinger-Hartree equation for non-Strichartz magnetic field was discussed. The global wellposedness and semiclassical limit of the Pauli-Poisson equation was discussed in \cite{moller2023poisson}.\\

\subsection{$N$-body Pauli equation}

Nonlinear 1-body PDE like the Schrödinger-Poisson equation arise as the mean field limit of linear $N$-body equations with interaction between the particles like the $N$-body Schrödinger equation with Coulomb interaction. A quantum system consisting of a large number $N$ of interacting particles is described by an $N$-body wave function
\begin{equation}
   \psi^{\hbar}_N =  \psi^{\hbar}_N(x_1,\dots ,x_N),
\end{equation}
where $x_j \in \mathbb{R}^3$. 
The wave function is normalized in $L^2(\mathbb{R}^{3N})$, i.e.
\begin{equation}
    \int_{\mathbb{R}^{3N}} |\psi^{\hbar}_N(x_1,\dots,x_N)|^2 \dd x_1 \cdots \dd x_N = 1,
\end{equation}
and satisfies the linear $N$-body non-relativistic Schrödinger equation
\begin{equation}
    i\hbar \partial_t \psi^{\hbar}_N = H_N \psi^{\hbar}_N
    \label{eq:n body schrödinger},
\end{equation}
where $H_N$ is the $N$-body Hamiltonian given by
\begin{equation}
    H_N = - \frac{\hbar^2}{2m} \sum_{j=1}^{N} \Delta_{x_j} + \frac{1}{N} \sum_{j<k}^N V(|x_j-x_k|)
    \label{eq:n body hamiltonian},
\end{equation}
where $V$ is some interaction potential. For the Coulomb interaction one has $V(x) \simeq |x|^{-1}$ and for the "contact interaction" $V(x) = \delta(x)$ which results in the Gross-Pitaevskii equation.  For large $N$, equation \eqref{eq:n body schrödinger} becomes impossible to solve numerically. Therefore it is imperative to approximate linear $N$-body equations by (systems of) nonlinear $1$-body equations. The following diagram represents  the asymptotic links between $N$-body linear and $1$-body nonlinear equations.\\

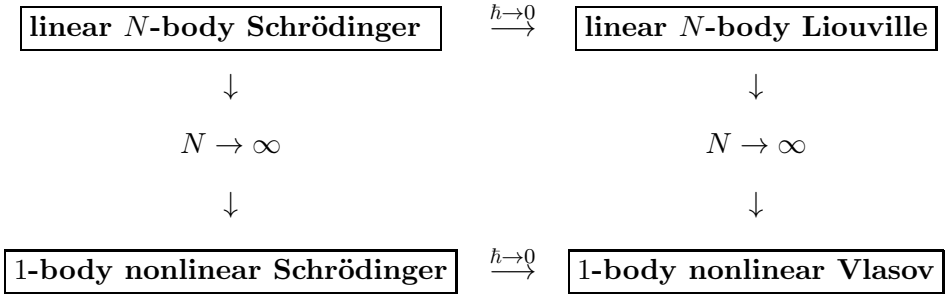
\begin{tabular}{ccc}
   {\fbox{\bf linear $N$-body Schrödinger  }}
    & {$\stackrel{\hbar \rightarrow 0}{\longrightarrow}$}
    & {\fbox{\begin{varwidth}{\textwidth}\centering\bf linear $N$-body Liouville \end{varwidth}}} \\[3mm]

  {$\downarrow$}
    &
    & {$\downarrow$} \\[3mm]
  {${N \rightarrow \infty}$}
    &
    & {${N \rightarrow \infty}$} \\[3mm]
  $\downarrow$
    &
    & $\downarrow$ \\[3mm]
  {\fbox{\bf $1$-body nonlinear Schrödinger}}
    & {$\stackrel{\hbar \rightarrow 0}{\longrightarrow} $} 
    & {\fbox{\begin{varwidth}{\textwidth}\centering\bf $1$-body nonlinear Vlasov\end{varwidth}}}
    \\
\end{tabular}\\
\bigskip

The Hartree ansatz for boson condensate, i.e. particles with symmetric $N$-body wave function, is to assume that the inital data are factorized with the same wave function for all bosons,
\begin{equation}
    \psi^{\hbar}_N(x_1,\dots ,x_N,t=0) = \prod_{j=1}^{N} \psi^{\hbar}_I(x_j)
\end{equation}
which produces a symmetric wave function. This  is valid for a pure state of a boson ensemble (i.e. if the system of bosons is in a condensed state). Note that the general Hartree ansatz for bosons would use different orbitals $\psi^{\hbar}_{I,j}$.

For fermions (i.e. for antisymmetric wave functions) a different ansatz has to be chosen. The \emph{Hartree-Fock ansatz} consists of taking initial fermionic, i.e. antisymmetric $N$-body wave functions $\psi^{\hbar}_{N,I} \in L^2_{\text{as}} (\mathbb{R}^{3N})$ (the subspace of $L^2$ consisting of antisymmetric (w.r.t. permutation of the arguments) wave functions) which give rise to a $N$-body Schrödinger evolution 
\begin{equation}
\psi_N^{\hbar} (x_1,\dots,x_N,t)= \exp(-iH_N t/\hbar) \psi^{\hbar}_{N,I} (x_1,\dots,x_N)
\end{equation}
where $H_N$ is the $N$-body Schrödinger Hamiltonian \eqref{eq:n body hamiltonian}. The associated initial (pure state) one particle reduced density matrix $\rho^{\hbar,(1)}_{N,I}$ should be close in the trace norm to the initial one particle reduced density matrix $\rho^{\hbar}_{\text{S},I}$ of the Slater determinant $\psi^{\hbar}_{\text{S}}$, where $\psi^{\hbar}_{\text{S}}$ is defined by
\begin{equation}
\psi^{\hbar}_{\text{S}}(x_1,\dots, x_N,t) = \frac{1}{\sqrt{N!}} \det(\psi^{\hbar}_i(x_j,t)),
\end{equation}
where $\{\psi^{\hbar}_j\}_{j=1}^{N}$ is an orthonormal system in $L^2(\mathbb{R}^3)$. The Slater determinant is a particular choice of an antisymmetric wave function. Then $\rho^{\hbar,(1)}_{N,I}$ should satisfy
\begin{equation}
    \tr(\rho^{\hbar,(1)}_{N,I}-\rho^{\hbar}_{\text{S},I}) \leq C,
\end{equation}
uniformly in $N$. The (pure state) time evolution $\rho^{\hbar}_{\text{S}}(t)$ of $\rho^{\hbar}_{\text{S},I}=\rho^{\hbar}_{\text{S}}(0)$ 
is given by
\begin{equation}
    \rho^{\hbar}_{\text{S}}(t) = \sum_{j=1}^N \ket{\psi^{\hbar}_j(t)} \bra{\psi^{\hbar}_j(t)},
\end{equation} 
and satisfies the \emph{time dependent Hartree-Fock (TDHF) equation}
\begin{equation}
    i\hbar \partial_t \rho^{\hbar}_{\text{S}}(t) = \left[-\frac{\hbar^2}{2}\Delta + \frac{1}{|x|} \ast \rho^{\hbar}_{\text{S},\text{diag}}(t) - X,\rho^{\hbar}_{\text{S}}(t)\right],
\end{equation}
where 
\begin{equation}
\rho^{\hbar}_{\text{S},\text{diag}}(x) = \frac{1}{N}\rho^{\hbar}_{\text{S}}(x,x)
\end{equation} is the density of $\rho^{\hbar}_{\text{S}}$ and $X$ denotes the exchange term with integral kernel
\begin{equation}
    X(x,y) = \frac{1}{N} \frac{1}{|x-y|}\rho^{\hbar}_{\text{S}}(x,y).
\end{equation} 
It is then expected that the time evolution  $\rho^{\hbar,(1)}_N(t)$ of the initial one particle reduced density matrix $\rho^{\hbar,(1)}_{N,I}$ should remain close to $\rho^{\hbar}_{\text{S}}(t)$ and their distance in trace norm should vanish in the limit $N\rightarrow \infty$. \\

The Pauli-Poisson equation should  arise as the mean field limit of the \textbf{$N$-body Pauli equation} given by 
\begin{equation}
    i\hbar \partial_t \Psi^{\hbar}_N = H^{\text{P}}_N \Psi^{\hbar}_N
    \label{eq:n body pauli},
\end{equation}
where $H^{\text{P}}_N$ is the linear $N$-body Pauli Hamiltonian with Coulomb interaction given by
\begin{equation}
    H^{\text{P}}_N = - \frac{1}{2m} \sum_{j=1}^{N} (\hbar\nabla_{x_j}-i\frac{q}{c}A(x_j))^2 - \frac{\hbar q}{2cm} (\sigma \cdot B(x_j)) + \frac{1}{N} \sum_{j<k}^N \frac{1}{|x_j-x_k|}
\label{eq:n body pauli hamiltonian},
\end{equation}
with initial data
\begin{equation}
    \Psi^{\hbar}_{N,I} = (\Psi_{I}^{\hbar})^{\otimes N} \in ((L^2(\mathbb{R}^3))^2)^{\otimes N} \cong (L^2(\mathbb{R}^{3N}))^2.
\end{equation}
The following diagram should hold for the Pauli equation:\\

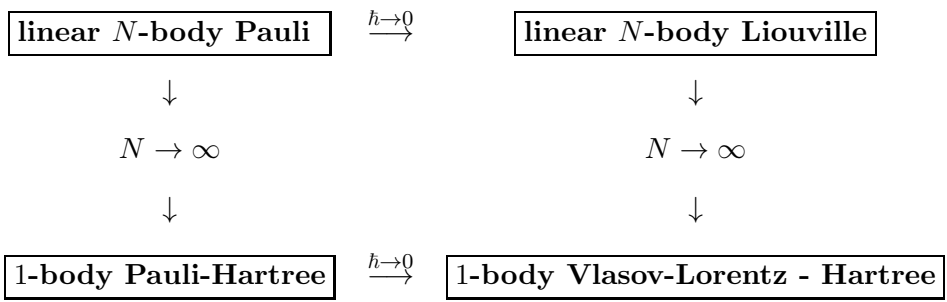
\begin{tabular}{ccc}
   {\fbox{\bf linear $N$-body Pauli  }}
    & {$\stackrel{\hbar \rightarrow 0}{\longrightarrow}$}
    & {\fbox{\begin{varwidth}{\textwidth}\centering\bf linear $N$-body Liouville \end{varwidth}}} \\[3mm]

  {$\downarrow$}
    &
    & {$\downarrow$} \\[3mm]
  {${N \rightarrow \infty}$}
    &
    & {${N \rightarrow \infty}$} \\[3mm]
  $\downarrow$
    &
    & $\downarrow$ \\[3mm]
  {\fbox{\bf $1$-body Pauli-Hartree}}
    & {$\stackrel{\hbar \rightarrow 0}{\longrightarrow} $} 
    & {\fbox{\begin{varwidth}{\textwidth}\centering\bf $1$-body Vlasov-Lorentz - Hartree\end{varwidth}}}%
    \\
\end{tabular}\\
\bigskip

Since the Pauli equation holds for fermions the Pauli exclusion principle implies that the Hartree ansatz as for the bosonic $N$-body magnetic Schrödinger equation is in fact not accurate. The correct ansatz is Hartree-Fock ansatz. However in practice the Hartree interaction is sufficient for numerics since the exchange term $X$ is small in most situations. In fact the \emph{Schrödinger-Poisson-$X\alpha$ equation} was proposed in \cite{mauser2001schrodinger} and studied numerically in \cite{bao2003effective}. The exchange term $X$ is replaced by a power nonlinearity $|\psi^{\hbar}|^{2/3} \psi^{\hbar}$ 
, based on a an approximation of the exchange term due to Slater \cite{slater1951simplification}.

\section{Asymptotic analysis}

In this section we emphasize the dependence on $\hbar$ and $N$ and use a scaling where $m=c=q=1$. The dependence on $c$ can be omitted since we only deal with the semiclassical and mean field limits and not with the non-relativistic limit.

\subsection{Semiclassical limit}

Mixed states in quantum mechanics represent a statistical ensemble of possibles states and are the fundamental object of quantum mechanics since a pure state is a special case of a mixed state. The mixed state formulation is necessary from a technical point of view when dealing with the semiclassical limit of the Schrödinger-Poisson and Pauli-Poisson equations since uniform $L^2$ estimates for the Wigner transform are only possible in a mixed state formulation. A mixed state is represented by the density matrix which is defined as follows.

Let $\{\Psi^{\hbar}_j\}_{j\in \mathbb{N}}$, $\Psi^{\hbar}_j = (\Psi_{j,1}^{\hbar},\Psi_{j,2}^{\hbar})^T$ be an orthonormal system in $(L^2(\mathbb{R}^3))^2$. We define the density matrix $\rho^{\hbar}$ and the matrix valued density matrix $R^{\hbar}$ as
\begin{align} \label{eq:Def_rho}
    \rho^{\hbar}(x,y,t) &:= \sum_{j=1}^{\infty} \lambda^{\hbar}_j \left(\Psi_{j,1}^{\hbar} (x,t) {\Psi_{j,1}^{\hbar}(y,t)^*} +\Psi_{j,2}^{\hbar} (x,t) {\Psi_{j,2}^{\hbar}(y,t)^*}\right)  , \\ R^{\hbar}(x,y,t) &:= \sum_{j=1}^{\infty} \lambda_j^{\hbar} \Psi_j^{\hbar}(x,t)\otimes \overline{\Psi_j^{\hbar}(y,t)},
\end{align}
where  $\lambda = \{\lambda_j^{\hbar}\}_{j\in \mathbb{N}}$ is a normally convergent series such that $\lambda_j^{\hbar} \geq 0$ and $\sum_j \lambda_j^{\hbar} = 1$. If  there is a $k$ such that $\lambda_j^{\hbar} = 1$ for $j=k$ and $\lambda_j^{\hbar} = 0$, otherwise it represents a \emph{mixed state}. The density matrix $\rho^{\hbar}$ can be considered as the kernel of a Hilbert-Schmidt, hermitian, positive and trace class operator $\varrho^{\hbar}$ on $L^2(\mathbb{R}^3)$, called \emph{density operator}. The diagonal of $\rho^{\hbar}(x,y)$ corresponds to the particle density and is defined by
\begin{align}
    \rho^{\hbar}_{\text{diag}}(x) &:= \rho^{\hbar}(x,x) = \sum_{j=1}^{\infty} \lambda_j^{\hbar} |\Psi_j^{\hbar}(x)|^2 \in L^1(\mathbb{R}^3_x).
    \label{eq:rho_diag} \\ R^{\hbar}_{\text{diag}}(x) &:= R^{\hbar}(x,x).
\end{align}
The time evolution of $\rho^{\hbar}$ is given by the von Neumann equation:
\begin{equation}
    i\hbar \frac{\partial \rho^{\hbar}}{\partial t} = [H,\rho^{\hbar}].
    \label{eq:von_Neumann_Liouville}
\end{equation}
The Wigner transform $f^{\hbar}(x,\xi,t)$ (resp. Wigner matrix $F^{\hbar}$) of $\rho^{\hbar}$ (resp. $R^{\hbar}$) is defined as (cf. \cite{gerard1997homogenization})
\begin{align}
    f^{\hbar}(x,\xi,t) &:= \frac{1}{(2\pi \hbar)^3} \int_{\mathbb{R}^3} e^{-i\xi\cdot y} \rho^{\hbar}(x+\frac{\hbar y}{2}, x-\frac{\hbar y}{2},t) \dd y,
    \label{eq:WT_rho} \\
       F^{\hbar}(x,\xi,t) &:= \frac{1}{(2\pi\hbar)^3} \int_{\mathbb{R}_y^3} e^{-i\xi \cdot y} R^{\hbar}(x+\frac{\hbar y}{2}, x-\frac{\hbar y}{2},t) \dd y.
    \label{eq:wigner_matrix}
\end{align}
Note that $f^{\hbar}=\Tr(F^{\hbar})$ and $\rho^{\hbar} = \Tr(R^{\hbar})$ where $\Tr$ denotes the $2\times 2$ matrix trace. A simple calculation shows that
\begin{align}
\label{eq:density_wigner}
    \rho^{\hbar}_{\text{diag}}(x) = \int_{\mathbb{R}^3_{\xi}} f^{\hbar}(x,\xi) \dd \xi, && R^{\hbar}_{\text{diag}}(x) = \int_{\mathbb{R}^3_{\xi}} F^{\hbar}(x,\xi) \dd \xi.
\end{align}
The \emph{mixed state Pauli-Poisson equation} is given by
\begin{align}
    i\hbar\partial_t \Psi_j^{\hbar} &= -\frac{1}{2}(\hbar \nabla-iA)^2\Psi_j^{\hbar} + V^{\hbar} \Psi_j^{\hbar} -\frac{1}{2} \hbar (\sigma \cdot B) \Psi_j^{\hbar}, \label{eq:PP_Pauli_mixed}\\
    -\Delta V^{\hbar} &= \sum_{j=1}^{\infty} \lambda_j^{\hbar} |\Psi_j^{\hbar}|^2 = \rho^{\hbar}_{\text{diag}}, \\
    \Psi_j^{\hbar}(x,0) &= \Psi^{\hbar}_{j,I}(x) \in (L^2(\mathbb{R}^3))^2.
    \label{eq:PP_data_mixed}
\end{align}
where the mixed state Pauli current density is given by
\begin{equation}
    J^{\hbar}(\Psi^{\hbar},A) = \sum_{j=1}^{\infty} \lambda_j^{\hbar}\left[\Im(\overline{\Psi_j^{\hbar}}({\hbar}\nabla -{i}A)\Psi_j^{\hbar}) -{\hbar}\nabla \times (\overline{\Psi_j^{\hbar}} \sigma \Psi_j^{\hbar})\right], \label{eq:PPW_current_mixed}
\end{equation}
Rewriting \eqref{eq:PP_Pauli_mixed}-\eqref{eq:PP_data_mixed} in the density matrix formulation using the von Neumann equation and taking its Wigner transform one obtains the \textbf{Pauli-Wigner-Poisson equation} for $F^{\hbar}$,
\begin{align}
\begin{split}
    \partial_t F^{\hbar} + \xi \cdot \nabla_x F^{\hbar} - \mathcal{F}_y[\beta[A]]\ast_{\xi} \nabla_x F^{\hbar} - \theta[A] (\xi F^{\hbar}) + \frac{1}{2}\theta[|A|^2]F^{\hbar}& \\ - \frac{\hbar}{2}\theta[\sigma \cdot B]F^{\hbar} +\theta[V^{\hbar}]F^{\hbar}&= 0, \label{eq:pauli_wigner}
\end{split}\\
-\Delta V^{\hbar} = \rho_{\text{diag}}^{\hbar} &= \int_{\mathbb{R}^3_{\xi}} f^{\hbar} \dd \xi, \\
F^{\hbar}(x,\xi,0) &= F^{\hbar}_I(x,\xi),
\label{eq:pauli wigner data}
\end{align}
where $\theta[\cdot]$ is the pseudo-differential operator defined by
\begin{equation}
    (\theta[\cdot]\Phi^{\hbar})(x,\xi,t) := \frac{1}{(2\pi)^3}\int_{\mathbb{R}^6} \delta[\cdot](x,y,t)\Phi^{\hbar}(x,\eta,t) e^{-i(\xi-\eta)\cdot y} \dd \eta \dd y. 
    \label{eq:PDO}
\end{equation}
where
\begin{equation}
    \beta[g] := \frac{1}{2}(g(x+\frac{\hbar y}{2})+g(x-\frac{\hbar y}{2})),
    \label{eq:beta}
\end{equation}
and 
\begin{equation}
    \delta[g] := \frac{i}{\hbar}(g(x+\frac{\hbar y}{2})-g(x-\frac{\hbar y}{2})).
    \label{eq:delta}
\end{equation}
\begin{assumption}
\label{thm:remark_pure_states}
Let $R^{\hbar}$ or $\rho^{\hbar}$ be a matrix valued density matrix or density matrix defined by an orthonormal system $\{\Psi_j^{\hbar}\}\subset (L^2(\mathbb{R}^3))^2$ and occupation probabilities $\lambda^{\hbar}_j \in [0,1]$. We assume that
\begin{align}
    \lambda_j^{\hbar} \geq 0, \quad \sum_{j=1}^{\infty} \lambda^{\hbar}_j = 1 ,
\end{align}
\begin{equation}
    \frac{1}{\hbar^3}\sum_{j=1}^{\infty} (\lambda^{\hbar}_j)^2 = \frac{1}{\hbar^3} \|\lambda^{\hbar}\|_2^2\leq C. \label{eq:weight_conditionC}
\end{equation}
Since \eqref{eq:weight_conditionC} implies that the sequence $\{\lambda^{\hbar}\}$ depends on $\hbar$ the reason for the superscript becomes apparent. This assumption implies uniform $L^2$ bounds in $\hbar$ for the Wigner transform.
\end{assumption}
We have the following theorem from \cite{moller2023poisson}.
\begin{theorem}
\label{thm:main}
Let $\{\Psi^{\hbar}_j\}_{j\in \mathbb{N}}\in C(\mathbb{R}_t,(L^2(\mathbb{R}_x^3))^2)$ be a solution of the mixed state Pauli-Poisson equation \eqref{eq:PP_Pauli_mixed}-\eqref{eq:PP_data_mixed} with associated matrix valued density matrix $R^{\hbar}$ such that the occuptation probabilities satisfy Assumption \ref{thm:remark_pure_states}. Let $F^{\hbar}$ be the associated Wigner matrix solving the Pauli-Wigner equation \eqref{eq:pauli_wigner} with initial data $F^{\hbar}_I(x,p) = F^{\hbar}(x,p,0)$. Assume that $F^{\hbar}_I$ converges up to a subsequence in $\mathcal{S}'(\mathbb{R}^3_x \times \mathbb{R}^3_p)^{2\times 2}$ to a nonnegative matrix-valued Radon measure $F_I$.
\begin{enumerate}[label=(\roman*)]
    \item \label{prop:C1} Let $A,V\in C^1(\mathbb{R}^3)$ such that $B :=\nabla \times A\in C(\mathbb{R}^3)$. 
Then $F^{\hbar}$ converges weakly* up to a subsequence in $(\mathcal{S}')^{2\times 2}$ to $F\in C_b(\mathbb{R}_t,\mathcal{M}^{2\times 2}_{w*})$ such that $F$ solves the Vlasov-Poisson equation with Lorentz force
\begin{equation}
     \partial_t F + p \cdot \nabla_x F +(-\nabla_x V + p\times B)\cdot \nabla_p F = 0,
     \label{eq:limit vlasov}
\end{equation}
in $(\mathcal{D}')^{2\times 2}$ verifying the initial condition
\begin{equation}
    F(x,p,0) = F_I(x,p) \quad \text{in }\mathbb{R}_x^3\times \mathbb{R}_p^3.
\end{equation}

\item \label{thm:semiclassical_limit_nonlinear}
Let $V^{\hbar}$ be given by $-\Delta V^{\hbar}= \rho^{\hbar}_{\text{\emph{diag}}}$ and suppose $A \in W^{1,\frac{7}{2}}(\mathbb{R}^3)$. 
Moreover, suppose that $\{F_I^{\hbar}\}$ is a bounded sequence in $(L^2(\mathbb{R}^6))^{2\times 2}$ and that the initial energy is bounded independently of $\hbar$. Then $F^{\hbar}$ converges weakly* up to a subsequence in $L^{\infty}(I,L^2(\mathbb{R}_x^3 \times \mathbb{R}_{p}^3)^{2\times 2})$ to 
\begin{equation*}
    F \in C_b(\mathbb{R}_t,\mathcal{M}^{2\times 2}_{w*})\cap L^{\infty}(I,L^1\cap L^2(\mathbb{R}_x^3 \times \mathbb{R}^3_{p})^{2\times 2})
\end{equation*}
such that $F$ solves
\begin{equation}
     \partial_t F + p \cdot \nabla_x F +(-\nabla_x V + p\times B))\cdot\nabla_p F = 0,
     \label{eq:limit vlasov poisson vlasov}
\end{equation}
in $(\mathcal{D}')^{2\times 2}$ and
\begin{equation}
    -\Delta V = \rho_{\text{\emph{diag}}}(x), \quad \rho_{\text{\emph{diag}}}(x) = \int_{\mathbb{R}^3_{\xi}} f(x,p) \dd p
    \label{eq:limit vlasov poisson poisson}
\end{equation}
where $f=\Tr(F)$, verifying the initial condition
\begin{equation}
    F(x,p,0) = F_I(x,p) \quad \text{in }\mathbb{R}_x^3\times \mathbb{R}_p^3.
    \label{eq:limit vlasov poisson data}
\end{equation}

\item \label{thm_Pauli_current}
Let $A \in W^{1,\frac{7}{2}}(\mathbb{R}^3)$. The mixed state Pauli current density $J^{\hbar}$ defined by
\begin{equation}
    J^{\hbar} = \sum_{j=1}^{\infty} \lambda_j^{\hbar} \left[ \Im(\overline{\Psi_j^{\hbar}}(\hbar \nabla -iA)\Psi_j^{\hbar}) - \hbar \nabla \times (\overline{\Psi_j^{\hbar}}\sigma \Psi_j^{\hbar}) \right]
\end{equation}
converges in $\mathcal{D}'$ to
\begin{equation}
    J = \int_{\mathbb{R}^3_p} pf \dd p.
\end{equation}
\end{enumerate}
\end{theorem}

\subsection{Mean field limit}

 For bounded interaction potential the bosonic $N$-body Schrödinger equation \eqref{eq:n body schrödinger} was shown in \cite{bardos2000weak, spohn1980kinetic} to converge to the Hartree equation
\begin{equation*}
    i\hbar\partial_t \psi^{\hbar} =  -\frac{\hbar^2}{2} \psi^{\hbar} + (V\ast |\psi^{\hbar}|^2) \psi^{\hbar}
\end{equation*}
This was extended in \cite{bardos2002derivation, erdos2001derivation} to the Coulomb potential
\begin{equation}
    V(x) = \frac{\lambda}{|x|}
\end{equation}
which implies the convergence of the three-dimensional $N$-body Schrödinger equation with Coulomb interaction to the Schrödinger-Poisson equation. \\

The convergence of the fermionic $N$-body Schrödinger equation to the TDHF equation was shown for bounded, symmetric binary interaction potentials $V$ (boundedness excludes the Coulomb potential) in \cite{bardos2003mean}. The problem of the convergence of the fermionic $N$-body Schrödinger equation with Coulomb interaction to the Hartree-Fock equation is hard due to the singular nature of the Coulomb potential. The Hartree-Fock dynamics for Coulomb interaction were proved in \cite{porta2017mean} for the scaling $\hbar = N^{-1/3}$ which links the mean field limit with the semiclassical limit, in \cite{petrat2017hartree} for a different scaling linking potential and kinetic energy and in \cite{frohlich2011microscopic} for the same scaling as in \cite{bardos2003mean}. It is shown that the fermionic $N$-body Schrödinger equation with Coulomb potential is approximated by the Hartree-Fock equation in the sense that the time evolutions stay close in the trace norm. This result holds for $\rho^{\hbar}_N$ representing a pure state, i.e. $\rho^{\hbar}_N$ is given by an orthogonal projection on the $N$-dimensional subspace spanned by antisymmetric wave functions. Notice that this is at odds with the semiclassical limit of the Schrödinger-Poisson equation in $\mathbb{R}^3$ to the Vlasov-Poisson equation \cite{lions1993mesures,markowich1993classical} and the limit of the Pauli-Poisson equation to the Vlasov-Poisson equation with Lorentz force in \cite{moller2023poisson} where only mixed states are allowed since the occupation probabilities have to satisfy conditon \eqref{eq:weight_conditionC} in order for the Wigner transform to be bounded uniformly in $L^2$. A recent result for the Hartree-Fock dynamics of fermionic mixed states is \cite{benedikter2016mean}, however it does not include Coulomb interaction. Moreover, in \cite{porta2017mean} the assumptions on the initial data for $\rho_N^{\hbar}$ are restrictive in the sense that one needs control over the commutator $[x,\rho_N^{\hbar}]$ for which the authors of \cite{porta2017mean} did not find non-trivial sufficient conditions. \\

For the bosonic magnetic Schrödinger equation with Coulomb interaction the $N$-body Hamiltonian is given by
\begin{equation}
    H^{\text{m}}_N = - \frac{1}{2} \sum_{j=1}^{N} (\hbar\nabla_{x_j}-iA(x_j))^2 + \frac{1}{N} \sum_{j<k}^N \frac{1}{|x_j-x_k|}
\end{equation}
The $N$-body wave function $\psi^{\hbar}_N$ satisfies the $N$-body magnetic Schrödinger equation
\begin{equation}
    i\hbar \partial_t \psi^{\hbar}_N = H^{\text{m}}_N \psi^{\hbar}_N
    \label{eq:n body msp}
\end{equation}
with initial data
\begin{equation}
    \psi^{\hbar}_{N,I} = (\psi_{I}^{\hbar})^{\otimes N} \in L^2(\mathbb{R}^{3N})
\end{equation}
It was shown by Lührmann \cite{luhrmann2012mean} that the linear $N$-body magnetic Schrödinger equation with Coulomb interaction converges to the magnetic Schrödinger-Hartree equation in the limit $N\rightarrow \infty$ for \emph{pure states} and for $\hbar$ \emph{fixed}. 

\begin{theorem}
    Let $A\in C^{\infty}$ such that
    \begin{align*}
        \|\partial^{\alpha} B(x)\| \leq C_{\alpha} \frac{1}{(1+|x|)^{-(1+\epsilon)}} && \|\partial^{\alpha} A(x)\| \leq C_{\alpha}
    \end{align*}
    for all $|\alpha|\geq 1$, $x\in \mathbb{R}^3$ and let $\psi^{\hbar}_{N,I}:= (\psi_I^{\hbar})^{\otimes N} \in H^{1}_A(\mathbb{R}^3N)$ be initial data to the $N$-body magnetic Schrödinger equation \eqref{eq:n body msp} such that $\|\psi^{\hbar}_{N,I}\|_2 = 1$. 
    Let $\rho^{\hbar,(k)}_N$ be the $k$-particle marginal density 
where $\rho^{\hbar}_N = \ket{\psi^{\hbar}_N}\bra{\psi^{\hbar}_N}$ and let $\psi^{\hbar}$ be the solution to the magnetic Schrödinger-Hartree equation \eqref{eq:MSP_Schrödinger} corresponding to the initial data $\psi^{\hbar}_I$. Then there exists a constant $C>0$ such that for $k\in \mathbb{N}$ and $t\in \mathbb{R}$,
        \begin{equation}
            \tr (\rho^{\hbar,(k)}_N - \ket{\psi^{\hbar}} \bra{\psi^{\hbar}}^{\otimes k}) \leq K \sqrt{8}\sqrt{\frac{k}{N}}e^{Ct}
        \end{equation}
        for all $k\leq N$. In particular, $\rho^{\hbar,(k)}_N$ converges in trace to $\ket{\psi^{\hbar}} \bra{\psi^{\hbar}}^{\otimes k}$ as $N\rightarrow \infty$.
\end{theorem}

\section{Wellposedness of Pauli-Poisson}

In this section we omit all superscripts since we do not consider asymptotics. We have the following global wellposedness result for the mixed state Pauli-Poisson equation \eqref{eq:PP_Pauli_mixed}-\eqref{eq:PP_data_mixed} from \cite{moller2023poisson}. Here,  $\underline{\Psi} := \{\Psi_j\}_{j\in \mathbb{N}}$ and $\underline{\Psi}_{I} = \{\Psi_{j,I}\}_{j\in \mathbb{N}}$. The energy space $\mathcal{H}^1(\mathbb{R}^3)$ is defined as \begin{equation*}
    \mathcal{H}^1(\mathbb{R}^3) := \{\underline{\Psi}\in \mathbf{L}^2 \colon (\nabla-iA)\underline{\Psi} \in \mathbf{L}^2, (\sigma \cdot B)_+^{1/2}\underline{\Psi} \in \mathbf{L}^2\} 
\end{equation*}
with associated norm
\begin{equation*}
    \|\underline{\Psi}\|^2_{\mathcal{H}^1} :=  \|(\nabla-iA)\underline{\Psi}\|^2_2 + \|(\sigma \cdot B)_+^{1/2}\underline{\Psi}\|^2_2 + \|\underline{\Psi}\|_2^2.
\end{equation*}

\begin{theorem}
\label{thm:PP_gwp}
Let $A\in L^2_{\text{\emph{loc}}}(\mathbb{R}^3)$, $|B| \in L^{2}(\mathbb{R}^3)$. For any $\Psi_I \in \mathcal{H}^1(\mathbb{R}^3)$ there exists a unique solution to the initial value problem \eqref{eq:PP_Pauli_mixed}-\eqref{eq:PP_data_mixed} in $C(\mathbb{R},\mathcal{H}^1(\mathbb{R}^3))\cap C^1(\mathbb{R},\mathcal{H}^1(\mathbb{R}^3)^*)$. If $\underline{\Psi}_{n,I},\underline{\Psi}_I \in \mathcal{H}^1(\mathbb{R}^3)$ are initial data satisfying $ \underline{\Psi}_{n,I}\rightarrow\underline{\Psi}_I$ in $\mathcal{H}^1(\mathbb{R}^3)$ with corresponding unique  solutions $\underline{\Psi}_n\in C(\mathbb{R},\mathcal{H}^1)\cap C^1(\mathbb{R},\mathcal{H}^{1*})$ and  $\underline{\Psi}\in C(\mathbb{R},\mathcal{H}^1)\cap C^1(\mathbb{R},\mathcal{H}^{1*})$ then $\underline{\Psi}_n \rightarrow \underline{\Psi}$ in $L^{\infty}(\mathbb{R},\mathcal{H}^1(\mathbb{R}^3))$.
\end{theorem}The global wellposedness in the energy space for the magnetic Schrödinger equation with Hartree nonlinearity $W \ast |\Psi|^2$ (including $W(x) = |x|^{-1}$) for the pure state case, i.e. \eqref{eq:MSP_Schrödinger}-\eqref{eq:MSP_data}, was proven in \cite{michelangeli2015global}. The magnetic Laplacian defines a self-adjoint operator on $\mathcal{H}^1$. Then one shows that the Hartree nonlinearity is Lipschitz in the energy space and finally one uses energy conservation to extend the solution globally. In \cite{michelangeli2015global}, the magnetic potential $A$ is assumed to be in $L^2_{\text{loc}}(\mathbb{R}^3)$ which is sufficient for the magnetic Laplacian $\Delta_A$ to be self-adjoint on $L^2(\mathbb{R}^3)$ due to a theorem by Leinfelder and Simader, cf. \cite{leinfelder1981schrodinger}. 

In \cite{barbaroux2017well} wellposedness of the magnetic Schrödinger-Poisson equation is proved for mixed states but only for bounded magnetic fields. Global wellposedness in $H^2$ of the Schrödinger-Poisson equation without magnetic field for mixed states was obtained in \cite{brezzi1991three},\cite{illner1994global} and in $L^2$ in \cite{castella1997l2}. 

We have the follwing  straightforward generalization of 
Theorem \ref{thm:PP_gwp} to the Pauli-Hartree equation.
\begin{theorem}[Global wellposedness of Pauli-Hartree]
\label{thm:PH_gwp}
Under the assumptions of Theorem \ref{thm:PP_gwp} and assuming that $W$ is even, $W \in L^{r_1}(\mathbb{R}^3) + L^{\infty}(\mathbb{R}^3)$ for $3/2 \leq r_1 \leq \infty$ and $\nabla W \in L^{r_2} + L^{\infty}$ for $1 \leq r_2 \leq \infty$ the Pauli-Hartree equation is globally wellposed in $\mathcal{H}^1(\mathbb{R}^3)$.
\end{theorem}

\begin{remark}
The question arises whether the Pauli-Hartree equation can be posed in arbitrary space dimensions. The three dimensional magnetic field $B=\nabla \times A$ has to be replaced by its $d$-dimensional generalization $\nabla \wedge A$. In this case, following the result for the magnetic Schrödinger-Hartree equation \cite{michelangeli2015global}, the conditions for $W$ would be: $W$ even, $W \in L^{r_1}(\mathbb{R}^3) + L^{\infty}(\mathbb{R}^3)$ for $\max\{1,d/2\} \leq r_1 \leq \infty$ ($r_1 > 1$ if $d=2$) and $\nabla W \in L^{r_2} + L^{\infty}$ for $\max\{1,d/3\} \leq r_2 \leq \infty$.
\end{remark}

\section*{Acknowledgement}

We acknowledge support from the Austrian Science Fund (FWF) via the grants SFB F65 and W1450 and by the Vienna Science and Technology Fund (WWTF) project MA16-066 "SEQUEX".

\bibliographystyle{abbrv}
{\small\bibliography{pdemodels}}

\end{document}